\documentclass[aps,12pt,superscriptaddress,amsfonts,amssymb,amsmath]{revtex4-2}
\pdfoutput=1

\usepackage{graphicx}
\usepackage{epsfig}
\usepackage{makeidx}
\usepackage{xcolor}
\usepackage{amsmath}
\usepackage{bm}
\usepackage{amssymb}

\begin{document}

\title{
Do we need an alternative to local gauge coupling to electromagnetic fields?
} 

\author{F.\ Minotti \footnote{Email address: minotti@df.uba.ar}}
\affiliation{Universidad de Buenos Aires, Facultad de Ciencias Exactas y Naturales, Departamento de F\'{\i}sica, Buenos Aires, Argentina}
\affiliation{CONICET-Universidad de Buenos Aires, Instituto de F\'{\i}sica Interdisciplinaria y Aplicada (INFINA), Buenos Aires, Argentina}

\author{G.\ Modanese \footnote{Email address: giovanni.modanese@unibz.it}}
\affiliation{Free University of Bozen-Bolzano \\ Faculty of Engineering \\ I-39100 Bolzano, Italy}
\date{\today}

\linespread{0.9}

\begin{abstract}

The local gauge coupling through the recipe $\partial_\mu \psi \to \partial_\mu \psi + iqA_\mu \psi$, that works so well with Dirac spinors in QED and in the gauge theories of the Standard Model, has a peculiarity when applied to scalar fields: it generates in the Lagrangian a coupling term $J_\mu A^\mu$ in which $J_\mu$ does not coincide with the conserved N\"other current associated to the global gauge symmetry.   This is not an inconsistency, just a feature that appears when working out the locally gauge invariant action, and which ensures that the correct conserved current is the source of the gauge field. What would happen then if we were to assume for the scalar field the same coupling $J_\mu A^\mu$ through a conserved current which holds for spinor QED and classical electrodynamics? The consequence is that one is forced in that case to renounce to the principle of local gauge symmetry and must thus consider the electromagnetic (e.m.) field to be described by electrodynamic theories compatible with that lack of invariance, like the extended electrodynamics by Aharonov-Bohm.   No differences with the usual theory appear for fermion systems when strict local charge conservation applies. In particular, if we consider the non-relativistic quantum theory as the low-energy limit of the relativistic theory, we would expect no modifications of Schr\"odinger equation when applied to fermion systems. However, when scalar boson systems are considered, like Cooper pairs quasi-particles in superconductors, in the new formulation the e.m.\ fields include a source, additional to the usual conserved four-current, and, besides, the corresponding Schr\"odinger equation acquires a new term, proportional to $\mathbf{A}^2$, which   can lead to   observable consequences, like a sizable change in the estimate of the magnetic penetration depth in 
  certain  
superconductors, compatible with the experimental data.
In conclusion, the alternative coupling considered yields a viable effective model for bosonic condensed matter systems, while for Dirac fermions it reduces to standard QED. Soft photon factorization and KLN cancellations in scalar QED fail in this framework, therefore particle physics scattering is outside the scope.  

\end{abstract}

\maketitle

\section{Introduction}

Local gauge invariance (LGI) is a powerful principle that, in the case of interaction of the considered system with e.m. fields, is further justified by the LGI of Maxwell equations. It is well known that LGI is intrinsically linked to strict local charge conservation (LCC). However, Aharonov and Bohm \cite{aharonov1963further} posited that LCC could be invalid at some small enough scale, and consequently built up an electrodynamics compatible with this non-conservation, which reduces to Maxwell's when LCC holds. Although they favored scales as small as Planck's for the failure of LCC, there are arguments for the possibility of non-conservation happening even at molecular scales \cite{li2008definition,zhang2011first,walz2015local,cabra2018simulation}. 
With these considerations, it could be argued that, at least for e.m. interactions, LGI is rather a consequence of LCC than its cause, and, consequently, it would be of interest to explore alternatives applicable if LCC happens to fail at some significant scale. 
What could thus be the alternative principle to determine the interaction of a general system with the e.m. field? In the following we argue that a possibility is that the interaction of the system with the e.m. field is quantified by a Lagrangian density term of the general form $J^{\mu}A_{\mu}$, where $A_{\mu}$ is the four potential of the e.m. field, and $J^{\mu}$ the four current resulting from $\boldsymbol{global}$ gauge invariance of the full Lagrangian. 
It is shown below that LGI and the alternative principle result in the same matter and interaction terms in the action of a Dirac field, whereas for a boson field the resulting actions differ. In the case of the Dirac field the e.m. term could thus be either Maxwell´s or that of Aharonov-Bohm (AB). If Maxwell expression is used LCC and the usual QED equations result, while with the AB e.m. term the resulting equations coincide with those of QED only when LCC is imposed.  
For the case of boson systems the alternative principle is not equivalent to LGI, and thus incompatible with a Maxwell e.m. term, so that the AB term is required. 
In this way we can say that, as a principle applicable to both, fermion and boson systems, we should consider the $J^{\mu}A_{\mu}$ interaction term discussed above, in conjunction with an AB e.m. term in both cases.
In the following sections we develop with some detail these ideas, and explore possible consequences of the application of the alternative principle to superconductors, considered as charged boson systems.

\section{Aharonov-Bohm electrodynamics and gauge invariance}

In order to introduce the general ideas we consider a system having e.m. interactions with e.m. fields given in terms of the four potential $
A_{\mu }=\left( \frac{1}{c}\varphi ,-\mathbf{A}\right) $. Greek indices denote the four coordinates, with $x_{0}=ct$, while Latin indices indicate
the three spatial coordinates.

We consider that the system, matter plus e.m. fields, is described by an action of the generic form
\begin{equation*}
S=\int_{R}\mathcal{L}\left( \phi ,\partial _{\mu }\phi \right) d^{4}x
\end{equation*}
in which $\phi $ represents all the matter and e.m. fields, and $R$ is a
space-time region bounded by two space-like hypersurfaces.

The equations of motion and conserved magnitudes are determined from this
action using Hamilton principle and Noether theorem. Since we will be concerned
only with conserved magnitudes arising from internal symmetries, space-time related symmetries will not be considered in the application of
Noether theorem.

Variation of the fields results in the variation of the action
\begin{eqnarray*}
\delta S &=&\int_{R}\left[ \frac{\partial \mathcal{L}}{\partial \phi }\delta
\phi +\frac{\partial \mathcal{L}}{\partial \left( \partial _{\mu }\phi
\right) }\delta \left( \partial _{\mu }\phi \right) \right] d^{4}x \\
&=&\int_{R}\left[ \frac{\partial \mathcal{L}}{\partial \phi }-\partial _{\mu
}\frac{\partial \mathcal{L}}{\partial \left( \partial _{\mu }\phi \right) }
\right] \delta \phi d^{4}x+\int_{\partial R}\frac{\partial \mathcal{L}}{
\partial \left( \partial _{\mu }\phi \right) }\delta \phi d\sigma _{\mu },
\end{eqnarray*}
where $\partial R$ corresponds to the boundary of $R$.

The application of Hamilton principle thus results in the equations of motion for
the fields
\begin{equation}
\frac{\partial \mathcal{L}}{\partial \phi }-\partial _{\mu }\frac{\partial 
\mathcal{L}}{\partial \left( \partial _{\mu }\phi \right) }=0.
\label{eqsmot}
\end{equation}

The considered version of Noether theorem, on the other hand, results for the case in which a
particular variation of the fields, of the general form $\delta \phi =\varepsilon
\Phi $, with $\varepsilon $ an infinitesimal parameter, leaves the action
invariant: $\delta S=0$. Assuming that the given variation is that of fields
that satisfy the equations of motion, the action invariance means that, for
a generic $\partial R$, 
\begin{equation*}
\int_{\partial R}\frac{\partial \mathcal{L}}{\partial \left( \partial _{\mu
}\phi \right) }\Phi d\sigma _{\mu }=0,
\end{equation*}
or, equivalently, $\partial _{\mu }J^{\mu }=0$, with
\begin{equation}
J^{\mu }=\frac{\partial \mathcal{L}}{\partial \left( \partial _{\mu }\phi
\right) }\Phi .  \label{Jmu}
\end{equation}

Let us consider as a concrete example the case of QED, in which the matter
field is described by the Lagrangian density
\begin{equation}
\mathcal{L}_{M}=\frac{i\hbar }{2}\left( \overline{\psi }\gamma ^{\mu
}\partial _{\mu }\psi -\partial _{\mu }\overline{\psi }\gamma ^{\mu }\psi
\right) -mc^{2}\overline{\psi }\psi ,  \label{LM_QED}
\end{equation}
in terms of Dirac four spinor and matrices, where the overbar indicates the
Pauli adjoint. The gauge principle, or minimal coupling principle, tells us
that matter of electric charge $q$ interacting with the e.m. field is
described by the Lagrangian (\ref{LM_QED}) in which partial derivatives are
replaced by covariant ones:
\begin{equation*}
\partial _{\mu }\rightarrow \partial _{\mu }+i\frac{q}{\hbar }A_{\mu },
\end{equation*}
that is
\begin{equation*}
\mathcal{L}_{M\_int}=\frac{i\hbar }{2}\left[ \overline{\psi }\gamma ^{\mu
}\left( \partial _{\mu }+i\frac{q}{\hbar }A_{\mu }\right) \psi -\overline{
\left( \partial _{\mu }+i\frac{q}{\hbar }A_{\mu }\right) \psi }\gamma ^{\mu
}\psi \right] -mc^{2}\overline{\psi }\psi ,
\end{equation*}
which can be written as 
\begin{equation*}
\mathcal{L}_{M\_int}=\mathcal{L}_{M}+\mathcal{L}_{int},
\end{equation*}
with
\begin{equation*}
\mathcal{L}_{int}=-J^{\mu }A_{\mu },
\end{equation*}
in which 
\begin{equation}
J^{\mu }=q\overline{\psi }\gamma ^{\mu }\psi .  \label{JmuQED}
\end{equation}

The full Lagrangian density for matter interacting with e.m. fields is thus
\begin{equation}
\mathcal{L}=\mathcal{L}_{M}+\mathcal{L}_{int}+\mathcal{L}_{em}
\label{L_M_int_em}
\end{equation}
where $\mathcal{L}_{em}$ describes the e.m. part and depends only on e.m. fields.

We note that in this approach three fundamental properties are in some sense supporting each other: local gauge invariance, current conservation and physical interpretation of the term $J^\mu A_\mu$ as matter-field interaction energy. In fact, if in this term one makes a local gauge transformation $A_\mu \to A_\mu+\partial_\mu f$, after integrating by parts $J^\mu (\partial_\mu f)$ and using conservation $\partial_\mu J^\mu=0$ one can confirm that the interaction energy is gauge invariant and thus a well-defined physical quantity.

We further note that $J^{\mu }$ given by (\ref{JmuQED}) is the conserved
current, as given by Noether theorem, resulting from the global gauge
invariance of the action, corresponding to variations of the fields of the form
\begin{subequations}
\label{global}
\begin{eqnarray}
\delta \psi  &=&-i\frac{q}{\hbar }\Lambda \psi , \\
\delta \overline{\psi } &=&i\frac{q}{\hbar }\Lambda \overline{\psi }, \\
\delta A^{\mu } &=&\partial ^{\mu }\Lambda =0,
\end{eqnarray}
with $\Lambda $ an infinitesimal constant.

We thus see from this familiar example that in QED there are two possible approaches to determine the interaction of charged matter with the e.m. field:

(a) We can use the minimal coupling principle, as done above.
This is the standard technique in gauge theory.
Equivalently, the physically relevant source is defined by functional differentiation with respect to
the gauge field. This definition guarantees that the current entering the field equations is conserved,
thanks to local gauge invariance and Ward-Takahashi identities.

(b) We can alternatively postulate a Lagrangian like (\ref{L_M_int_em}), in
which $\mathcal{L}_{M}$ is the non-interacting matter field Lagrangian, $
\mathcal{L}_{em}$ describes the e.m. part, and the interaction term is $\mathcal{L}_{int}=-J^{\mu
}A_{\mu }$, with $J^{\mu }$ the conserved current resulting from the global
gauge invariance of the full action.

Note however that both approaches are not equivalent, for instance, for a
charged scalar matter field, with Lagrangian
\end{subequations}
\begin{equation}
\mathcal{L}_{scalar}=\hbar ^{2}\partial ^{\mu }\phi ^{\ast }\partial _{\mu
}\phi -\kappa ^{2}\phi ^{\ast }\phi . \label{Lmscalar}
\end{equation}

If we include its interaction with the e.m. field using the
principle of minimal coupling we obtain
\begin{equation*}
\mathcal{L}_{scalar\_int}=\hbar ^{2}\left( \partial ^{\mu }-i\frac{q}{\hbar }
A^{\mu }\right) \phi ^{\ast }\left( \partial _{\mu }+i\frac{q}{\hbar }A_{\mu
}\right) \phi -\kappa ^{2}\phi ^{\ast }\phi ,
\end{equation*}
or, in expanded form,
\begin{equation*}
\mathcal{L}_{scalar\_int}=\hbar ^{2}\partial ^{\mu }\phi ^{\ast }\partial
_{\mu }\phi -\kappa ^{2}\phi ^{\ast }\phi +q\left( i\phi \partial ^{\mu
}\phi ^{\ast }-i\phi ^{\ast }\partial ^{\mu }\phi +q\phi ^{\ast }\phi A^{\mu
}\right) A_{\mu }.
\end{equation*}

We can thus write $\mathcal{L}_{scalar\_int}=\mathcal{L}_{scalar}+\mathcal{L}
_{int}$, with $\mathcal{L}_{int}=-J^{\mu }A_{\mu }$, where
\begin{equation}
J^{\mu }=iq\hbar\left( \phi ^{\ast }\partial ^{\mu }\phi -\phi \partial ^{\mu
}\phi ^{\ast }\right) -q^{2}\phi ^{\ast }\phi A^{\mu }.  \label{JHuang}
\end{equation}
This, however, is not the conserved current associated to global gauge invariance of the full action. In fact, the action corresponding to $\mathcal{L}_{scalar\_int}$ is invariant for a global gauge transformation
\begin{eqnarray*}
\delta \phi &=&-i\frac{q}{\hbar }\Lambda \phi , \\
\delta \phi ^{\ast } &=&i\frac{q}{\hbar }\Lambda \phi ^{\ast }, \\
\delta A^{\mu } &=&\partial ^{\mu }\Lambda =0,
\end{eqnarray*}
and so a direct application of Noether theorem gives as the conserved current the expression
\begin{equation}
J_{conserved}^{\mu }=iq\hbar\left( \phi ^{\ast }\partial ^{\mu }\phi -\phi
\partial ^{\mu }\phi ^{\ast }\right) -2q^{2}\phi ^{\ast }\phi A^{\mu }.
\label{Jscalar_conserved}
\end{equation}

  As mentioned above, this is not an inconsistency because  the use of the generally non-conserved current (\ref{JHuang}) for the interaction Lagrangian results in a $\mathcal{L}_{scalar\_int}$
that is invariant for local gauge transformations, and so perfectly
consistent with an e.m. field described by the Maxwell Lagrangian
\begin{equation}
\mathcal{L}_{em}=-\frac{1}{4\mu _{0}}F^{\mu \nu }F_{\mu \nu }.
\label{Lem_Maxwell}
\end{equation}
 In fact,  by variation of the four potential $A_{\mu}$ in the full Lagrangian the usual Maxwell equations are thus obtained, with the conserved source (\ref{Jscalar_conserved}).

On the other hand, if we were to take as a principle that the interaction
part of the Lagrangian is given in terms of the conserved current (\ref
{Jscalar_conserved}), the resulting $\mathcal{L}_{scalar\_int}$ would not be
invariant for local gauge transformations, so that $\mathcal{L}_{em}$ could
not be (\ref{Lem_Maxwell}). Indeed, it is readily seen that if we use (\ref
{Lem_Maxwell}) in this case, the resulting Maxwell equations have as source the four current
\begin{equation*}
J^{\mu }=iq\hbar\left( \phi ^{\ast }\partial ^{\mu }\phi -\phi \partial ^{\mu
}\phi ^{\ast }\right) -4q^{2}\phi ^{\ast }\phi A^{\mu },
\end{equation*}
which is not conserved in general.

Note that the contribution of a term proportional to $A^\mu A_\mu$ to the matter-field interaction energy is not gauge-invariant, but cannot in general be reduced to zero through a gauge transformation $A_\mu \to A_\mu+\partial_\mu f$ such that $(A_\mu+\partial_\mu f)(A^\mu+\partial^\mu f)=0$. As a simple example, consider a magnetostatic limit with stationary currents but no free charges and no electric field. In this case the condition on $f$ becomes $\mathbf{A}^2+2\mathbf{A}\cdot \nabla f+(\nabla f)^2=0$, whose solution is $\nabla f=-\mathbf{A}$, whence $\mathbf{B}=0$, incompatible with the hypothesized presence of currents.

The origin of the difference between the case of a scalar field and the QED case can be traced to the matter Lagrangian (\ref{Lmscalar}) not being linear in the derivatives of the matter field. We can thus expect that both principles for including the interaction of matter with the e.m. field are not equivalent for matter fields with such kind of Lagrangian. Since Lagrangians that are not linear in the derivatives of the matter field arise in the fundamental description of bosons, we will refer to bosons as the fields for which both principles are not equivalent.   

A possible way to make the $J^{\mu}A_{\mu}$ principle for bosons compatible with the e.m. field equations is to take AB Lagrangian for the e.m. part:
\begin{equation}
\mathcal{L}_{em}^{AB}=-\frac{1}{4\mu _{0}}F^{\mu \nu }F_{\mu \nu }-\frac{1}{
2\mu _{0}}\left( \partial ^{\mu }A_{\mu }\right) ^{2},  \label{LemAB}
\end{equation}
which is compatible with non-conserved sources. In Sect.\ \ref{sch-bos} we shall implement this idea for the case of a non-relativistic boson field.

\subsection{Implications of the use of AB Lagrangian}

The important question is now what would be the consequences of using the AB Lagrangian for the e.m. part, instead of Maxwell's, in any Lagrangian describing matter interacting with e.m. fields. 

An interesting point about the AB Lagrangian is that it can be recast as
\begin{equation*}
\mathcal{L}_{em}^{AB}=-\frac{1}{2\mu _{0}}\left( \partial _{\mu }A_{\nu
}\right) \left( \partial ^{\mu }A^{\nu }\right) +\frac{1}{2\mu _{0}}\partial
_{\nu }\left( A^{\mu }\partial _{\mu }A^{\nu }-A^{\nu }\partial _{\mu
}A^{\mu }\right) ,
\end{equation*}
so that leaving out the surface term it has the simplest possible expression for a spin 1 massless field.

Furthermore the AB Lagrangian is not invariant for local gauge transformations, but it is invariant for a global gauge transformation of the form (\ref{global}). This means that if it is used as the e.m. part in, for instance, the QED Lagrangian (\ref{L_M_int_em}), Noether theorem tells us that the current (\ref{JmuQED}) is conserved. This means
that the resulting non-homogeneous equation for the e.m. field
\begin{equation*}
\partial ^{\nu }\partial _{\nu }A^{\mu }=\mu _{0}J^{\mu },
\end{equation*}
satisfies
\begin{equation*}
\partial ^{\nu }\partial _{\nu }\left( \partial _{\mu }A^{\mu }\right) =0,
\end{equation*}
whose solution, for any distribution of localized sources, is $\partial _{\mu
}A^{\mu }=0$. 
This means that the AB Lagrangian coincides with Maxwell's and
no difference is observed in the phenomenology between both approaches. 

In this way, any phenomenology of AB electrodynamics, novel to that of
Maxwell, is suppressed in this case by local charge conservation. Furthermore, this
local charge conservation is the consequence of the global gauge invariance
of the AB Lagrangian itself. This apparently means that there is no need of AB electrodynamics for fermions.

A possible loophole out of this situation is given by Noether theorem
itself, in which the conservation of the current requires that the fields satisfy the
dynamic equations (\ref{eqsmot}) that result from Hamilton principle. However,
a quantum measurement or, more generally, a quantum collapse or localization
process does not apparently follow those equations \cite{ghirardi1986_PhysRevD.34.470,ghirardi_1990_PhysRevA.42.78,zurek2003,schlosshauer2005,galapon2009}, and so it is possible that during this kind of process charge is not locally conserved. 

For systems in which the principles of
minimal coupling and of an interaction term with the conserved current are not equivalent, an additional possibility exists. In this case, as shown above, the dynamic equations for the
e.m. field have generally a locally non-conserved source, so that a non-Maxwell e.m.\
part of the Lagrangian, as AB's, is required. It is important to note that in this case there could be strict local conservation of charge, the difference being that an additional source, that acts as a non-conserved current, is added to the e.m. field equations.

With these considerations it would result that fermion systems, either described by QED or its low-energy non-relativistic approximations show an AB phenomenology only when, for instance, quantum collapse processes, leading to violation of LCC, are present. 

For boson systems the same possibility occurs if the minimal coupling principle operates, but with additional possibilities if, instead, the $J^{\mu}A_{\mu}$ principle were valid. 
In a series of previous works \cite{Modanese2017480,Modanese2017MPLB,modanese2017electromagnetic,modanese2018time,Minotti-Modanese-Symmetry2021,minotti2021quantum,minotti2022electromagnetic,minotti2023aharonov,EPJC2023,minotti_modanese_math13050892} we have explored the possibility of the
violation of LCC by quantum processes, like non-local interactions, quantum
uncertainty, tunneling, etc. and the ensuing phenomenology.

In the following we consider the alternative possibility in which the $J^{\mu }A_{\mu }$ principle is assumed valid, and study some of its consequences for boson systems. 

In particular, we explore the description given in terms of Schr\"{o}dinger-like equations, and its application to superconductors. 

\subsection{Schr\"{o}dinger-like equation for bosons without minimal coupling}
\label{sch-bos}

Let us consider a boson system described by Schr\"{o}dinger equation, whose non-interacting matter Lagrangian can be written as: 
\begin{equation}
\mathcal{L}_{M}=\frac{i\hbar c}{2}\left( \psi ^{\ast }\partial _{0}\psi
-\psi \partial _{0}\psi ^{\ast }\right) -\frac{\hbar ^{2}}{2m}\nabla \psi
\cdot \nabla \psi ^{\ast },  \label{LMSchroedinger}
\end{equation}

For a global gauge variation, of the form 
\begin{eqnarray*}
\delta \psi &=&-i\frac{q}{\hbar }\Lambda \psi , \\
\delta \psi ^{\ast } &=&i\frac{q}{\hbar }\Lambda \psi ^{\ast }, \\
\delta A^{\mu } &=&\partial ^{\mu }\Lambda =0,
\end{eqnarray*}
the matter action $S_{M}=\int \mathcal{L}
_{M}d^{4}x$ is clearly invariant, and its contribution to the conserved
current is, using Noether theorem for $S_{M}$,
\begin{eqnarray*}
J_{M}^{0} &=&qc\psi ^{\ast }\psi , \\
J_{M}^{j} &=&\frac{iq\hbar }{2m}\left( \psi \partial _{j}\psi ^{\ast }-\psi
^{\ast }\partial _{j}\psi \right) .
\end{eqnarray*}

Since $S_{em}=\int \mathcal{L}_{em}d^{4}x$ does not depend on the matter
field, there is no contribution from it to the conserved current.

In this way, the interaction part of the action contains a term of the form: 
$-\int J_{M}^{\mu }A_{\mu }d^{4}x$, which in turn has a contribution to the
conserved current, obtained applying Noether theorem to this part, given by 
\begin{eqnarray*}
J_{int}^{0} &=&0, \\
J_{int}^{j} &=&-\frac{q^{2}}{m}\psi ^{\ast }\psi A_{j}.
\end{eqnarray*}
Note that this expression verifies that its contribution to the action is invariant for a global gauge transformation, as it must.

Since this part of the current does not contain derivatives of the matter
field there is no additional contribution to the total conserved current.

In this way we finally have for the conserved current the usual expression 
\begin{subequations}
\label{conservedJ}
\begin{eqnarray}
J^{0} &=&J_{M}^{0}+J_{int}^{0}=qc\psi ^{\ast }\psi , \\
J^{j} &=&J_{M}^{j}+J_{int}^{j}=\frac{iq\hbar }{2m}\left( \psi \partial
_{j}\psi ^{\ast }-\psi ^{\ast }\partial _{j}\psi \right) -\frac{q^{2}}{m}
\psi ^{\ast }\psi A_{j}.
\end{eqnarray}

It is readily checked that variation of the matter field in the resulting total action gives the Schr\"{o}dinger-like equation 
\end{subequations}
\begin{equation}
i\hbar \frac{\partial \psi }{\partial t}=\frac{1}{2m}\left( -i\hbar \nabla -q
\mathbf{A}\right) ^{2}\psi +q\varphi \psi +\frac{q^{2}}{2m}\left\vert 
\mathbf{A}\right\vert ^{2}\psi .
\label{sch-like}
\end{equation}
Note that the last term is missing in the standard version of the equation, as obtained using the minimal coupling principle. Furthermore, from this equation we can readily verify that the conserved current is precisely (\ref{conservedJ}).

When we derive the equations for the e.m. field from the variation of the
action due to variations of the four potential $A_{\mu }$, we must take into account
that $J^{\mu }$ depends on $A_{\mu }$: 
\begin{equation*}
\delta S=\int_{R}\left[ \frac{\partial \mathcal{L}_{em}}{\partial A_{\mu }}
\delta A_{\mu }+\frac{\partial \mathcal{L}_{em}}{\partial \left( \partial
_{\nu }A_{\mu }\right) }\delta \left( \partial _{\nu }A_{\mu }\right)
-\left( J^{\mu }+A_{\nu }\frac{\delta J^{\nu }}{\delta A_{\mu }}\right)
\delta A_{\mu }\right] d^{4}x,
\end{equation*}
which by Hamilton principle results in:
\begin{eqnarray*}
\frac{\partial \mathcal{L}_{em}}{\partial A_{0}}-\partial _{\nu }\frac{
\partial \mathcal{L}_{em}}{\partial \left( \partial _{\nu }A_{0}\right) }
&=&J^{0}, \\
\partial _{\nu }\frac{\partial \mathcal{L}_{em}}{\partial \left( \partial
_{\nu }A_{j}\right) }-\frac{\partial \mathcal{L}_{em}}{\partial A_{j}}
&=&J^{j}-\frac{q^{2}}{m}\psi ^{\ast }\psi A_{j}.
\end{eqnarray*}

Finally note that one can include other potentials $V\left( x^{\mu }\right) $
by adding terms to the matter Lagrangian (\ref{LMSchroedinger}), like, for
instance, $V\left( x^{\mu }\right) \psi \psi ^{\ast }$, or, more generally,
a function of $\psi \psi ^{\ast }=\left\vert \psi \right\vert ^{2}$, $
F\left( x^{\mu },\left\vert \psi \right\vert ^{2}\right) $, without altering
the previous considerations and the expression of the four current, but only
modifying the Schr\"{o}dinger-like equation, to include terms like $V\psi $,
or $\partial F/\partial \left\vert \psi \right\vert ^{2}\psi =$ $V\left(
x^{\mu },\left\vert \psi \right\vert ^{2}\right) \psi $.

With all these considerations, if we take $\mathcal{L}_{em}$ to be AB's (\ref
{LemAB}), we obtain the full set of equations 
\begin{subequations}
\label{superfull}
\begin{eqnarray}
i\hbar \frac{\partial \psi }{\partial t} &=&\frac{1}{2m}\left( -i\hbar
\nabla -q\mathbf{A}\right) ^{2}\psi +q\varphi \psi +V\left( x^{\mu
},\left\vert \psi \right\vert ^{2}\right) \psi +\frac{q^{2}}{2m}\left\vert 
\mathbf{A}\right\vert ^{2}\psi ,  \label{finalpsi} \\
\frac{1}{c^{2}}\frac{\partial ^{2}\varphi }{\partial t^{2}} &=&\nabla
^{2}\varphi +\frac{\rho }{\varepsilon _{0}},  \label{finalphi} \\
\frac{1}{c^{2}}\frac{\partial ^{2}\mathbf{A}}{\partial t^{2}} &=&\nabla ^{2}
\mathbf{A+}\mu _{0}\mathbf{j}_{c}-\frac{\mu _{0}q^{2}}{m}\left\vert \psi
\right\vert ^{2}\mathbf{A},  \label{finalA}
\end{eqnarray}
where 
\end{subequations}
\begin{eqnarray}
\rho &=&q\left\vert \psi \right\vert ^{2},  \label{rho_cons} \\
\mathbf{j}_{c} &=&\frac{iq\hbar }{2m}\left( \psi \nabla \psi ^{\ast }-\psi
^{\ast }\nabla \psi \right) -\frac{q^{2}}{m}\left\vert \psi \right\vert ^{2}
\mathbf{A}.  \label{j_cons}
\end{eqnarray}
In Eq. (\ref{finalA}) we have split the e.m. source into the current that satisfies local conservation
\begin{equation*}
\frac{\partial \rho }{\partial t}+\nabla \cdot \mathbf{j}_{c}=0,
\end{equation*}
and the generally non-conserved additional source $-\frac{q^{2}}{m}\left\vert \psi
\right\vert ^{2}\mathbf{A}$.

It is important to recall that in AB electrodynamics invariance under
arbitrary local gauge transformations is lost (although global gauge invariance still
applies), and that the potentials must be determined by the last two of
equations (\ref{superfull}) in their present form.

\subsection{Application to superconductors}
\label{applicationSC}

\subsubsection{Modified London theory}
\label{modified-london}

We consider a theory of superconductivity in which the matter Lagrangian (\ref{finalpsi}) describes the dynamics of the wave function of the condensate. We
will explicitly denote the number density of Cooper pairs $
n_{s}=\left\vert \psi \right\vert ^{2}$, while the rest of ions is simply
described as a neutralizing fixed background with charge $-q$ and uniform
number density $n_{0}$, so that the electric charge density is
\begin{equation*}
\rho =qn_{s}-qn_{0}.
\end{equation*}

If we consider quasistatic conditions ($c^{-2}\left( \partial /\partial
t\right) ^{2}\ll \nabla ^{2}$) we have from Eq. (\ref{finalA})
\begin{equation}
\nabla ^{2}\mathbf{A}-\lambda _{L}^{-2}\mathbf{A}=-\mu _{0}\mathbf{j}_{c},
\label{LondonA}
\end{equation}
where $\lambda _{L}$ is the London penetration depth
\begin{equation*}
\lambda _{L}=\left( \frac{m}{\mu _{0}q^{2}n_{s}}\right) ^{1/2}.
\end{equation*}

We can now proceed in the usual way followed in London theory to determine
the distribution of current that minimizes the free energy.

In AB electrodynamics the e.m. energy density is given by \cite{minotti2021quantum}
\begin{equation*}
u_{em}=\frac{1}{\mu _{0}}\left[ \frac{\left\vert \mathbf{E}\right\vert ^{2}}{
2c^{2}}+\frac{\left\vert \mathbf{B}\right\vert ^{2}}{2}+\frac{\varphi }{c^{2}
}\frac{\partial S}{\partial t}-\mathbf{A}\cdot \nabla S-\frac{S^{2}}{2}
\right] ,
\end{equation*}
where
\begin{eqnarray*}
\mathbf{E} &=&-\nabla \varphi -\frac{\partial \mathbf{A}}{\partial t}, \\
\mathbf{B} &=&\nabla \times \mathbf{A}, \\
S &=&\frac{1}{c^{2}}\frac{\partial \varphi }{\partial t}+\nabla \cdot 
\mathbf{A}.
\end{eqnarray*}

The e.m. contribution to the free energy is thus given by $U_{em}=\int
u_{em}dV$, where the integral is extended to a given volume. For fixed
values of the fields on the volume surface the variation of $U_{em}$ can be
readily written as
\begin{eqnarray*}
\delta U_{em} &=&\frac{1}{\mu _{0}}\int \left[ -\frac{1}{c^{2}}\nabla
^{2}\varphi \delta \varphi -\nabla ^{2}\mathbf{A}\cdot \delta \mathbf{A}+
\frac{1}{c^{2}}\frac{\partial \mathbf{A}}{\partial t}\cdot \frac{\partial
\delta \mathbf{A}}{\partial t}\right. \\
&&\left. +\frac{\partial ^{2}\varphi }{\partial t^{2}}\frac{\delta \varphi }{
c^{4}}+\frac{\varphi }{c^{4}}\frac{\partial ^{2}\delta \varphi }{\partial
t^{2}}-\frac{1}{c^{4}}\frac{\partial \varphi }{\partial t}\frac{\partial
\delta \varphi }{\partial t}\right] dV.
\end{eqnarray*}

For quasistatic conditions the above expression reduces to
\begin{equation*}
\delta U_{em}=-\frac{1}{\mu _{0}}\int \left[ \frac{1}{c^{2}}\nabla
^{2}\varphi \delta \varphi +\nabla ^{2}\mathbf{A}\cdot \delta \mathbf{A}
\right] dV.
\end{equation*}

Furthermore, since the conserved current density $\mathbf{j}_{c}$ determines
the drift velocity $\mathbf{v}_{D}$ of the supercurrent: $\mathbf{j}
_{c}=qn_{s}\mathbf{v}_{D}$, we can express the kinetic energy density as
\begin{equation*}
u_{k}=\frac{m}{2n_{s}q^{2}}\left\vert \mathbf{j}_{c}\right\vert ^{2}=\frac{1
}{2}\mu _{0}\lambda _{L}^{2}\left\vert \mathbf{j}_{c}\right\vert ^{2},
\end{equation*}
so that the kinetic energy contribution to the free energy is $U_{k}=\int
u_{k}dV$.

We thus have
\begin{equation*}
\delta U_{k}=\int \mu _{0}\lambda _{L}^{2}\mathbf{j}_{c}\cdot \delta \mathbf{
j}_{c}dV=\int \mathbf{j}_{c}\cdot \left( \delta \mathbf{A}-\lambda
_{L}^{2}\nabla ^{2}\delta \mathbf{A}\right) dV,
\end{equation*}
where (\ref{LondonA}) was used in the last equality.

In this way, the condition $\delta U_{em}+\delta U_{k}=0$, leads to
\begin{eqnarray*}
\nabla ^{2}\varphi &=&0, \\
\frac{1}{\mu _{0}}\int \nabla ^{2}\mathbf{A}\cdot \delta \mathbf{A}dV
&=&\int \mathbf{j}_{c}\cdot \left( \delta \mathbf{A}-\lambda _{L}^{2}\nabla
^{2}\delta \mathbf{A}\right) dV.
\end{eqnarray*}

From the first relation and the quasistatic version of (\ref{finalphi}) we
see that $n_{s}=n_{0}$, so that $\lambda _{L}$ can be considered as uniform
in the volume of interest, which allows to obtain, from the second relation
\begin{equation}
\nabla ^{2}\mathbf{A}=\mu _{0}\left( \mathbf{j}_{c}-\lambda _{L}^{2}\nabla
^{2}\mathbf{j}_{c}\right) .  \label{LondonB}
\end{equation}

Equations (\ref{LondonA}) and (\ref{LondonB}) allow to determine the vector
potential and the conserved current for the equilibrium state in this simple
model.

Note further that, since $n_{s}=n_{0}$, we must also have $\nabla \cdot 
\mathbf{j}_{c}=0$. From Eqs. (\ref{LondonA}) and (\ref{LondonB}) this
implies that $\nabla \cdot \mathbf{A}=0$, so that the total current is also
conserved.

Note that the condition $\nabla \cdot \mathbf{A}=0$ is imposed by the theory, and is not the result of a gauge choice, which is not allowed in AB electrodynamics.

We now consider a slab geometry with $x$ the coordinate of variation inside the
material, with the superconductor state valid for $x>0$, while the normal
state applies for all $x<0$. The vector potential has a component in the
direction $\mathbf{e}_{y}$ of value $A_{0}$ at $x=0$. Equations (\ref
{LondonA}) and (\ref{LondonB}) have then as solution inside the
superconductor
\begin{eqnarray*}
\mathbf{A} &=&A_{0}\exp \left( -\frac{x}{\lambda }\right) \mathbf{e}_{y}, \\
\mathbf{j}_{c} &=&j_{0}\exp \left( -\frac{x}{\lambda }\right) \mathbf{e}_{y},
\end{eqnarray*}
which must satisfy
\begin{eqnarray*}
\frac{A_{0}}{\lambda _{L}^{2}}\left( 1-\xi ^{2}\right) &=&\mu _{0}j_{0}, \\
\frac{A_{0}}{\lambda _{L}^{2}}\frac{\xi ^{2}}{1-\xi ^{2}} &=&\mu _{0}j_{0},
\end{eqnarray*}
where $\xi =\lambda _{L}/\lambda $.

The compatibility condition
\begin{equation*}
1-\xi ^{2}=\frac{\xi ^{2}}{1-\xi ^{2}}
\end{equation*}
results in the possible solutions 
\begin{equation*}
\xi ^{2}=\frac{3\pm \sqrt{5}}{2},
\end{equation*}
with
\begin{equation*}
\mathbf{j}_{c}=\frac{1}{\mu _{0}\lambda _{L}^{2}}\left( 1-\xi ^{2}\right) 
\mathbf{A},
\end{equation*}
and total current
\begin{equation*}
\mathbf{J}=\mathbf{j}_{c}-\frac{1}{\mu _{0}\lambda _{L}^{2}}\mathbf{A}=-
\frac{\xi ^{2}}{\mu _{0}\lambda _{L}^{2}}\mathbf{A}.
\end{equation*}

The corresponding magnetic and kinetic energy per unit area in the
superconductor region is given by
\begin{eqnarray*}
U &=&\frac{1}{2}\int_{0}^{\infty }\left( \frac{\left\vert \mathbf{A}
\right\vert ^{2}}{\mu _{0}\lambda ^{2}}+\mu _{0}\lambda _{L}^{2}\left\vert 
\mathbf{j}_{c}\right\vert ^{2}\right) dx \\
&=&\frac{1}{2\mu _{0}\lambda _{L}^{2}}\left[ \xi ^{2}+\left( 1-\xi
^{2}\right) ^{2}\right] \int_{0}^{\infty }\left\vert \mathbf{A}\right\vert
^{2}dx \\
&=&\frac{A_{0}^{2}}{4\mu _{0}\lambda _{L}}\left[ \frac{\xi ^{2}+\left( 1-\xi
^{2}\right) ^{2}}{\xi }\right] .
\end{eqnarray*}

For fixed $A_{0}$ and $\lambda _{L}$, of the two possible solutions the
minimum of $U$ results for $\xi ^{2}=\frac{3-\sqrt{5}}{2}$, so that 
\begin{eqnarray*}
\lambda &=&\sqrt{\frac{2}{3-\sqrt{5}}}\lambda _{L}\simeq 1.618\lambda _{L},
\\
\mathbf{J} &\simeq &0.382\mathbf{J}_{L},
\end{eqnarray*}
where $\mathbf{J}_{L}$ is the current in London theory:
\begin{equation*}
\mathbf{J}=-\frac{1}{\mu _{0}\lambda _{L}^{2}}\mathbf{A}.
\end{equation*}

On the other hand, if instead of referring to the value $A_{0}$ at $x=0$, we
use the value of the magnetic field, whose magnitude at that position is $
B_{0}=A_{0}/\lambda $, we obtain for the energy
\begin{equation*}
U=\frac{B_{0}^{2}\lambda _{L}}{4\mu _{0}}\left[ \frac{\xi ^{2}+\left( 1-\xi
^{2}\right) ^{2}}{\xi ^{3}}\right] ,
\end{equation*}
which, for fixed $B_{0}$ and $\lambda _{L}$, presents a minimum for $\xi
^{2}=\frac{3+\sqrt{5}}{2}$, resulting in 
\begin{eqnarray*}
\lambda &=&\sqrt{\frac{2}{3+\sqrt{5}}}\lambda _{L}\simeq 0.618\lambda _{L},
\\
\mathbf{J} &\simeq &2.618\mathbf{J}_{L},
\end{eqnarray*}

Since equilibrium requires that the normal region has a magnetic field,
parallel to the interface normal-superconductor, of value equal to the
critical magnetic field at the given temperature, $B_{0}$ should be fixed at
its critical value, which favors the choice $\lambda \simeq 0.618\lambda_{L} $ in a material in which normal and superconducting regions coexist in
equilibrium.

In these derivations, based on the minimization of the free energy, the kinetic energy contribution is to be considered as an approximation. We can alternatively derive a result for the modified $\lambda_{L}$ assuming instead the spatial uniformity of $n_s$, without recourse to the free energy, in the following way.

Start from the expression of the conserved current in terms of the wavefunction:
\begin{equation}
    \mathbf{j}_c=\frac{\hbar q}{2mi}\left( 
\bar{\psi}\nabla\psi-\psi\nabla\bar{\psi} \right)-\frac{q^2}{m}|\psi|^2\mathbf{A}
\end{equation}

When $|\psi|^2$ is constant, denoted with $n_s$, the term independent from $\mathbf{A}$ can be written as proportional to the gradient of the phase of $\psi$:
\begin{equation}
    \mathbf{j}_c=\frac{n_s\hbar e}{2mi}\nabla\varphi-\frac{n_s q^2}{m}\mathbf{A}
\end{equation}
It follows that
\begin{equation}
    \nabla\times\mathbf{j}_c=-\frac{n_s q^2}{m}\mathbf{B}
    \label{rotJ}
\end{equation}

This is of course exactly like in the traditional London theory. Now we use the two new equations for  $\mathbf{j}_c$ and $\mathbf{A}$, relations (\ref{j_cons}) and (\ref{LondonA}). We know that in the case of constant density they imply $\nabla\cdot\mathbf{A}=0$. It follows from vector identities that in this case one has
\begin{equation}
    \nabla\times\mathbf{B}=\nabla\times\nabla\times\mathbf{A}=-\nabla^2\mathbf{A}
\end{equation}
Therefore we can write
\begin{equation}
    \nabla\times\mathbf{B}=-\nabla^2\mathbf{A}=\mu_0\mathbf{j}_c-\lambda_L^{-2}\mathbf{A}
    \label{rotB}
\end{equation}
where $\lambda_L^{-2}=n\mu_0q^2/m$ is the usual expression of $\lambda_L$. In the second equality we used Equation (\ref{LondonA}). Note here the change with respect to the traditional London theory, in which one simply has, using the fourth Maxwell equation, that $\nabla\times\mathbf{B}=\mu_0\mathbf{j}_c$.

Now we take the curl of \eqref{rotB} and obtain
\begin{equation}
    \nabla\times\nabla\times\mathbf{B}=\mu_0\nabla\times\mathbf{j}_c-\lambda_L^{-2}\mathbf{B}
\end{equation}
from which by expanding the double curl (remembering the third Maxwell equation, which also holds in AB electrodynamics) and using the expression \eqref{rotJ} for $\nabla\times\mathbf{j}_c$:
\begin{equation}
    -\nabla^2\mathbf{B}=-\frac{n_sq^2\mu_0}{m} \mathbf{B}-\lambda_L^{-2}\mathbf{B}
\end{equation}
and finally
\begin{equation}
    \nabla^2\mathbf{B}=2\frac{1}{\lambda^2_L}\mathbf{B} =\frac{1}{\left(\frac{\lambda_L}{\sqrt{2}}\right)^2}\mathbf{B}
\end{equation}
The conclusion is that the magnetic penetration length is reduced by a factor $1/\sqrt{2}\simeq 0.707$ compared to $\lambda_L$, which is to be compared with the approximate previous result of a factor $\simeq 0.618$. 

\subsubsection{Estimation of the surface impedance}

In order to determine possible verifiable effects predicted by the present
theory we proceed to estimate the surface impedance of a superconductor. For this we
consider the general equations (\ref{finalphi}) and (\ref{finalA}) for the
potentials, with a constant number density $n_{s}$ of Cooper pairs.

In the Fourier representation $a\left( \mathbf{x},t\right) \rightarrow
a\left( \mathbf{k}\right) \exp \left[ \mathbf{k}\cdot \mathbf{x}-\omega
\left( \mathbf{k}\right) t\right] $ we thus have (space-time and Fourier
transformed fields are noted with the same symbol): 
\begin{eqnarray*}
\left( k^{2}-\frac{\omega ^{2}}{c^{2}}\right) \varphi &=&\frac{\mathbf{k}
\cdot \mathbf{j}_{c}}{\varepsilon _{0}\omega }, \\
\left( k^{2}-\frac{\omega ^{2}}{c^{2}}\right) \mathbf{A} &=&\mu _{0}\mathbf{
j }_{c}-\frac{\mathbf{A}}{\lambda _{L}^{2}},
\end{eqnarray*}
where charge conservation in Fourier representation, $-i\omega \rho +i 
\mathbf{k}\cdot \mathbf{j}_{c}=0$, was used to write the charge density.

We further consider the normal and superconducting components of the current
density, $\mathbf{j}_{c}=\mathbf{j}_{n}+\mathbf{j}_{s}$, where the normal
component is given by $\mathbf{j}_{n}=\sigma _{n}\mathbf{E}$, with $\sigma
_{n}$ the complex conductivity of the normal electrons.

For the superconducting component we consider a simple fluid model, $\mathbf{j}_{s}=qn_{s}\mathbf{v}_{D}$, where the undamped mean drift velocity is determined by
the acceleration due to only the electric force: $-i\omega m\mathbf{v}_{D}=q
\mathbf{E}$, so that 
\begin{equation*}
\mathbf{j}_{s}=i\frac{q^{2}n_{s}}{m\omega }\mathbf{E}=i\frac{\mathbf{E}}{\mu
_{0}\lambda _{L}^{2}\omega }.
\end{equation*}

Using further that $\mathbf{E}=i\omega \mathbf{A}-i\mathbf{k}\varphi $, we
finally obtain the equations for the potentials in the superconducting
medium: 
\begin{eqnarray*}
\left[ k^{2}-\frac{\omega ^{2}}{c^{2}}+\frac{k^{2}}{\varepsilon _{0}\omega }
\left( i\sigma _{n}-\frac{1}{\mu _{0}\lambda _{L}^{2}\omega }\right) \right]
\varphi  &=&\left( i\sigma _{n}-\frac{1}{\mu _{0}\lambda _{L}^{2}\omega }
\right) \frac{\mathbf{k}\cdot \mathbf{A}}{\varepsilon _{0}}, \\
\left[ k^{2}-\frac{\omega ^{2}}{c^{2}}-\mu _{0}\omega \left( i\sigma _{n}-
\frac{2}{\mu _{0}\lambda _{L}^{2}\omega }\right) \right] \mathbf{A} &=&-\mu
_{0}\left( i\sigma _{n}-\frac{1}{\mu _{0}\lambda _{L}^{2}\omega }\right) 
\mathbf{k}\varphi .
\end{eqnarray*}

In order to evaluate the surface impedance we need to determine the ratio of
the electric and magnetic field components, parallel to the surface, of a
transverse electromagnetic wave. For the transverse mode ($\mathbf{k}\cdot 
\mathbf{A}=0$) we obtain from the previous equations that $\varphi =0$, and 
\begin{equation}
k^{2}=\frac{\omega ^{2}}{c^{2}}+i\mu _{0}\sigma _{n}\omega -\frac{2}{\lambda
_{L}^{2}}.  \label{k2transverse}
\end{equation}

In this way, noting that for the considered mode one has $\mathbf{E}=i\omega 
\mathbf{A}$, $\mathbf{B}=i\mathbf{k}\times \mathbf{A}$, the surface
impedance is given as 
\begin{eqnarray*}
Z_{s} &=&\mu _{0}\frac{E_{\Vert }}{B_{\Vert }}=\frac{\mu _{0}\omega }{k} \\
&=&\frac{\mu _{0}c}{\sqrt{1+i\frac{\mu _{0}c^{2}\sigma _{n}}{\omega }-\frac{2c^{2}}{\lambda _{L}^{2}\omega ^{2}}}}.
\end{eqnarray*}

We note that the same derivation, but starting with Maxwell equations for
the potentials in Lorentz gauge, results in an expression similar to (\ref
{k2transverse}), with the only difference that the number 2 in the last term
is replaced by unity, so that the present theory differs from the usual one
by the previously found scaling of the penetration length, $\lambda
_{L}\rightarrow $ $\lambda _{L}/\sqrt{2}$.

\subsubsection{Modified Ginzburg-Landau (GL) theory}
\label{mod-gl}

In this section we explore an application of Aharonov-Bohm electrodynamics with reduced gauge invariance to the Ginzburg-Landau theory for superconductors. For this purpose we introduce a few non-trivial assumptions and approximations, therefore the application of the resulting theoretical model to specific superconductors will require further elaboration.

The idea is to consider the expression of the free energy proposed by GL and
modify it according to the previous ideas: the covariant derivatives are
replaced by ordinary partial derivatives plus the inclusion of the
interaction term $-\mathbf{j}_{c}\cdot \mathbf{A}$, instead of the
corresponding term $-\left( \mathbf{j}_{c}+\frac{q^{2}}{2m}\left\vert \psi
\right\vert ^{2}\mathbf{A}\right) \cdot \mathbf{A}$ that would result from
the use of covariant derivatives. Furthermore, the Maxwell expression of the
magnetic energy is replaced by the AB one. 
We thus obtain for the free energy
\begin{eqnarray}
F_{s} &=&F_{n0}+a\left\vert \psi \right\vert ^{2}+\frac{b}{2}\left\vert \psi
\right\vert ^{4}+\frac{1}{2m}\left\vert -i\hbar \nabla \psi \right\vert ^{2}
\notag \\
&&-\frac{iq\hbar }{2m}\left( \psi \nabla \psi ^{\ast }-\psi ^{\ast }\nabla
\psi \right) \cdot \mathbf{A}+\frac{q^{2}}{m}\left\vert \psi \right\vert
^{2}\left\vert \mathbf{A}\right\vert ^{2}  \notag \\
&&+\frac{1}{2\mu _{0}}\left[ \left\vert \nabla \times \mathbf{A}\right\vert
^{2}-2\mathbf{A}\cdot \nabla \left( \nabla \cdot \mathbf{A}\right) -\left(
\nabla \cdot \mathbf{A}\right) ^{2}\right] ,  \label{GLext}
\end{eqnarray}
where $F_{n0}$ is the free energy in the normal state in zero magnetic
field, and $a$ and $b$ temperature dependent constants.

In the applications, in which the superconductor is immersed in an
externally controlled magnetic field $\mathbf{H}_{ext}$ the Legendre
transform to the Gibbs free energy is used:
\begin{equation*}
G_{s}=F_{s}-\mathbf{B}\cdot \mathbf{H}_{ext}.
\end{equation*}

The condition of thermodynamic stability is that $\int $ $G_{s}d^{3}x$ be
stationary for arbitrary variations of $\psi $ and $\mathbf{A}$, which,
taking into account that $\nabla \times \mathbf{H}_{ext}=0$ in the volume of
variation, is readily seen to lead to the equations 
\begin{subequations}
\label{GLeqs}
\begin{eqnarray}
0 &=&\frac{1}{2m}\left( -i\hbar \nabla -q\mathbf{A}\right) ^{2}\psi +a\psi
+b\left\vert \psi \right\vert ^{2}\psi +\frac{q^{2}}{2m}\left\vert \mathbf{A}
\right\vert ^{2}\psi ,  \label{GL1} \\
\nabla ^{2}\mathbf{A} &=&-\mu _{0}\left[ \frac{iq\hbar }{2m}\left( \psi
\nabla \psi ^{\ast }-\psi ^{\ast }\nabla \psi \right) -\frac{2q^{2}}{m}
\left\vert \psi \right\vert ^{2}\mathbf{A}\right] .  \label{GL2}
\end{eqnarray} 
\end{subequations}

These equations are analogous to Eqs. (\ref{superfull}).

A first thing to note is that the expression of the conserved supercurrent (\ref{j_cons}) is the same as in the original GL theory, so that the
quantization of the magnetic flux is also valid in the modified theory without any change.

To study how the phenomenology of the original GL theory is modified, we consider a simple slab geometry with fields dependent only on the spatial
coordinate $z$, with the superconducting medium at $z\rightarrow \infty $,
while $z\rightarrow -\infty $ corresponds to the normal state. Assuming an
imposed magnetic field in the $x$ direction $\mathbf{e}_{x}$, we take a
vector potential of the form $\mathbf{A}=A\left( z\right) \mathbf{e}_{y}$.
The corresponding equations are then 
\begin{subequations}
\label{GLslab}
\begin{eqnarray}
\frac{\hbar ^{2}}{2m}\frac{d^{2}\psi }{dz^{2}} &=&a\psi +b\left\vert \psi
\right\vert ^{2}\psi +\frac{q^{2}}{m}A^{2}\psi ,  \label{GLslabpsi} \\
\frac{d^{2}A}{dz^{2}} &=&\frac{2\mu _{0}q^{2}}{m}\left\vert \psi \right\vert
^{2}A,  \label{GLslabA} \\
\psi \frac{d\psi ^{\ast }}{dz} &=&\psi ^{\ast }\frac{d\psi }{dz}.
\end{eqnarray}

The last relation indicates that the phase of $\psi $ does not depend on $z$
and thus does not affect the other equations, in which only the real
magnitude of $\psi $ enters. We thus can take $\psi $ as real.

The boundary conditions are 
\end{subequations}
\begin{eqnarray*}
\psi &=&0,\text{ }\frac{dA}{dz}=B_{c}\text{ for }z\rightarrow -\infty , \\
\psi &=&\psi _{\infty },\text{ }A=0\text{ for }z\rightarrow \infty .
\end{eqnarray*}

The surface energy is defined in the usual way: the difference between the
Gibbs free energy per unit area between the actual state and that in which
the whole sample is in the normal state at the critical field $B_{c}$ \cite{fetter2012quantum}
\begin{eqnarray*}
\sigma _{s} &=&\int_{-\infty }^{\infty }\left[ G_{s}\left( z\right) -G_{n0}+
\frac{B_{c}^{2}}{2\mu _{0}}\right] dz \\
&=&\int_{-\infty }^{\infty }\left[ a\psi ^{2}+\frac{b}{2}\psi ^{4}+\frac{
\hbar ^{2}}{2m}\left( \frac{d\psi }{dz}\right) ^{2}+\frac{q^{2}}{m}\psi
^{2}A^{2}+\frac{\left( B-B_{c}\right) ^{2}}{2\mu _{0}}\right] dz.
\end{eqnarray*}

Multiplication of the first of (\ref{GLslab}) by $\psi $ and integration by
parts gives 
\begin{equation*}
\int_{-\infty }^{\infty }\left[ a\psi ^{2}+b\psi ^{4}+\frac{\hbar ^{2}}{2m}
\left( \frac{d\psi }{dz}\right) ^{2}+\frac{q^{2}}{m}\psi ^{2}A^{2}\right]
dz=0,
\end{equation*}
so that we can write
\begin{equation*}
\sigma _{s}=\int_{-\infty }^{\infty }\left[ -\frac{b}{2}\psi ^{4}+\frac{
\left( B-B_{c}\right) ^{2}}{2\mu _{0}}\right] dz,
\end{equation*}
which has exactly the same expression as that corresponding to the original
GL theory. Since in zero field ($\mathbf{A}=0$) the modified and original GL
equations coincide we have also the relations
\begin{eqnarray*}
\frac{a^{2}}{b} &=&\frac{B_{c}^{2}}{\mu _{0}}, \\
\left\vert \psi _{\infty }\right\vert  &=&\sqrt{\frac{\left\vert
a\right\vert }{b}},
\end{eqnarray*}
which allow us to write
\begin{equation*}
\sigma _{s}=\frac{B_{c}^{2}}{2\mu _{0}}\delta ,
\end{equation*}
with
\begin{equation*}
\delta =\int_{-\infty }^{\infty }\left[ \left( 1-\frac{B}{B_{c}}\right)
^{2}-\left( \frac{\psi }{\psi _{\infty }}\right) ^{4}\right] dz.
\end{equation*}

In terms of the non-dimensional variables
\begin{eqnarray*}
\widetilde{\psi } &=&\frac{\psi }{\psi _{\infty }}, \\
\widetilde{A} &=&A\sqrt{\frac{q^{2}}{m\left\vert a\right\vert }}, \\
\varsigma  &=&z\frac{\sqrt{2}}{\lambda _{L}}, \\
\kappa  &=&\frac{\sqrt{m\left\vert a\right\vert }\lambda _{L}}{\hbar },
\end{eqnarray*}
with
\begin{equation*}
\lambda_{L}=\left( \frac{m}{\mu _{0}q^{2}\psi _{\infty }^{2}}\right) ^{1/2},
\end{equation*}
the Eqs. (\ref{GLslabpsi}) and (\ref{GLslabA}) are written as
\begin{eqnarray*}
\frac{d^{2}\widetilde{\psi }}{d\varsigma ^{2}} &=&\kappa ^{2}\left[ \left( 
\widetilde{A}^{2}-1\right) \widetilde{\psi }+\widetilde{\psi }^{3}\right] ,
\\
\frac{d^{2}\widetilde{A}}{d\varsigma ^{2}} &=&\widetilde{\psi }^{2}
\widetilde{A}.
\end{eqnarray*}
The corresponding boundary conditions are 
\begin{eqnarray*}
\widetilde{\psi } &=&0,\text{ }\frac{d\widetilde{A}}{d\varsigma }=\frac{1}{
\sqrt{2}}\text{ for }\varsigma \rightarrow -\infty , \\
\widetilde{\psi } &=&1,\text{ }\widetilde{A}=0\text{ for }\varsigma
\rightarrow \infty .
\end{eqnarray*}

These equations and boundary conditions have the same expressions than those
in GL theory, for the GL non-dimensional variables
\begin{eqnarray*}
\widetilde{A}_{GL} &=&\sqrt{2}\widetilde{A}, \\
\varsigma _{GL} &=&\frac{\varsigma}{\sqrt{2}} , \\
\kappa _{GL} &=&\sqrt{2}\kappa .
\end{eqnarray*}

We thus see that a similar phenomenology results, in which only the
characteristic parameters are scaled. The characteristic magnetic
penetration length is $\lambda _{L}/\sqrt{2}$ instead of $\lambda _{L}$, as for the modified London's theory, and
the GL parameter $\kappa _{GL}$ is replaced by $\kappa =\kappa _{GL}/\sqrt{2}$.

\subsubsection{Superconductor classification and critical fields}

To determine the critical magnetic field for a given superconductor as
predicted by the modified GL equation we essentially follow \cite{degennes}. We consider an infinite sample
immersed in a strong external magnetic field. This field destroys
superconductivity and is uniform in the sample. If the magnetic field
intensity is lowered to the level where few Cooper pairs appear we can
linearize the modified GL equation in $\psi $ and, besides, take the vector
potential as that determined by the external field only, since the
supercurrents are proportional to $\psi ^{2}$. We thus write
\begin{equation*}
0=\frac{1}{2m}\left( -i\hbar \nabla -q\mathbf{A}\right) ^{2}\psi +a\psi +
\frac{q^{2}}{2m}\left\vert \mathbf{A}\right\vert ^{2}\psi ,
\end{equation*}
which, in expanded form, and using cylindrical coordinates ($r,\phi ,z$),
with the external field $B_{0}$ in the $z$ direction, is written as
\begin{equation}
-\frac{\hbar ^{2}}{2m}\nabla ^{2}\psi +\frac{i\hbar qB_{0}}{2m}\frac{
\partial \psi }{\partial \phi }+\frac{q^{2}B_{0}^{2}}{4m}r^{2}\psi =-a\psi . \label{Schlike}
\end{equation}

For the solution of this Shr\"{o}dinger-like equation one can use separation of variables so that
\begin{equation*}
\psi \left( r,\phi ,z\right) =R\left( r\right) Z\left( z\right) \exp \left(
ip\phi \right) ,
\end{equation*}
with $p$ an integer or zero. The lowest energy level corresponds to $p=0$ and $
\partial /\partial z=0$, so that the equation in this case reduces to
\begin{equation*}
-\frac{\hbar ^{2}}{2mr}\frac{d}{dr}\left( r\frac{dR}{dr}\right) +\frac{
q^{2}B_{0}^{2}}{4m}r^{2}R=-aR.
\end{equation*}
This expression coincides with the Schr\"{o}dinger equation for the fundamental level
of a particle of charge $q$ and mass $m$ in a uniform magnetic field of
value $\sqrt{2}B_{0}$. The corresponding energy is given by $E=\hbar \omega
_{c}/2$, with the cyclotron angular frequency $\omega _{c}=\left\vert
q\right\vert \sqrt{2}B_{0}/m$. 
We thus have
\begin{equation}
-a=\left\vert a\right\vert =\frac{\hbar \left\vert q\right\vert B_{0}}{\sqrt{
2}m}.  \label{a_critical}
\end{equation}

Using the GL expressions of the penetration depth $\lambda $ and correlation length $\xi $:
\begin{eqnarray*}
\lambda ^{2} &=&\frac{m}{\mu _{0}q^{2}\left\vert \psi \right\vert ^{2}}, \\
\xi ^{2} &=&\frac{\hbar ^{2}}{2m\left\vert a\right\vert }=\frac{\hbar
^{2}\left\vert \psi \right\vert ^{2}\mu _{0}}{2mB_{c}^{2}},
\end{eqnarray*}
we can write relation (\ref{a_critical}) as 
\begin{equation*}
B_{0}=\frac{\lambda }{\xi }B_{c}=\kappa _{GL}B_{c}.
\end{equation*}

If $\kappa _{GL}<1$ when the strong field is lowered Copper pair formation
starts a $B_{c}$, and the superconductor is thus of type I. 
For $\kappa_{GL}>1$ pair formation starts above $B_{c}$, at $B_{c2}=\kappa B_{c}$, and
we have a type-II superconductor. 
Note that for a given $a$ the consideration of higher energy levels of the Schr\"{o}dinger-like equation (\ref{Schlike}) would result in lower values of the critical field, which are thus of no consequence.

We finally note that the same result can be obtained for the simple slab geometry discussed
above, noting that in
that case the original GL theory coincides with the modified one with the scaled $\kappa =\kappa _{GL}/\sqrt{2}$. In this way, as determined in the original theory, the surface energy changes sign at $\kappa =1/\sqrt{2}$, which corresponds to $
\kappa _{GL}=1$. 

The present theory thus predicts that the high critical field in type-II superconductors is determined by $B_{c2}=\kappa_{GL} B_{c}$. 

To study the low critical field we also follow \cite{degennes}. As mentioned above, fluxoid quantization as determined in the original GL theory is not modified by the present theory, because the expression of the conserved current is not changed. In this way, to determine the low critical field $H_{c1}$ in a type-II
superconductor we consider the case in which the sample is immersed in a
weak enough external field for the Meissner effect to hold. If the external
field is increased penetration of one quantum of fluxoid $\varphi _{0}=h/q$
eventually occurs. As a model of this situation we consider an extreme
type-II superconductor with $\xi \ll \lambda $,  and take the magnetic field
in the $z$ direction with the fluxoid $\varphi _{0}$ contained in a
cylindrical region of radius $\xi $. In the superconducting region $r>\xi $
we neglect the spatial variation of $\psi $ so that equation (\ref{GL2})
reduces to  
\begin{equation*}
\nabla ^{2}\mathbf{A}=\frac{2\mu _{0}q^{2}}{m}\left\vert \psi \right\vert
^{2}\mathbf{A}=2\lambda ^{-2}\mathbf{A},
\end{equation*}
whose solution for $\mathbf{A}=A\left( r\right) \mathbf{e}_{\phi }$
corresponding to a fluxoid $\varphi _{0}$ is  
\begin{equation*}
A=\frac{\varphi _{0}}{2\pi \xi K_{1}\left( \sqrt{2}\xi /\lambda \right) }
K_{1}\left( \sqrt{2}r/\lambda \right) ,
\end{equation*}
where $K_{1}$ is the modified Bessel function of order one. With this
solution, neglecting the contribution from the region $r<\xi $, due to the
condition $\xi \ll \lambda $, the difference between the Gibbs energies per
unit of $z,$ 
\begin{equation*}
\int_{\xi }^{\infty }\left( F_{s}-\mathbf{B}\cdot \mathbf{H}_{ext}\right)
2\pi rdr,
\end{equation*}
of the Meissner and of the one-vortex states is determined from (\ref{GLext}) to be proportional to 
\begin{equation*}
\frac{\varphi _{0}}{4\pi \sqrt{2}\lambda ^{2}}\frac{K_{0}\left( u\right) 
\left[ K_{2}\left( u\right) -K_{0}\left( u\right) \right] }{K_{1}\left(
u\right) }-\mu _{0}H_{ext}K_{1}\left( u\right) ,
\end{equation*}
where $u=\sqrt{2}\xi /\lambda $. 

In this way, both states have the same Gibbs energy, thus allowing the spontaneous transition between them, for a value of the external field given by
\begin{eqnarray}
\mu _{0}H_{c1} &=&\frac{\varphi _{0}}{4\pi \sqrt{2}\lambda ^{2}}\frac{
K_{0}\left( u\right) \left[ K_{2}\left( u\right) -K_{0}\left( u\right) 
\right] }{K_{1}^{2}\left( u\right) } \notag \\
&\simeq &\frac{\varphi _{0}\sqrt{2}}{4\pi \lambda ^{2}}\ln \left( \sqrt{2}
\frac{\lambda }{\xi }\right) ,\label{Hc1}
\end{eqnarray}
where the condition $\xi \ll \lambda $ was used in the second line. GL theory allows to derive a similar expression where both $\sqrt{2}$ appearing in the last line of (\ref{Hc1}) are replaced by one.

\subsection{Comparison with experimental data}

As we have seen in the previous sections, the London equation is slightly modified in the AB extended electrodynamics and in particular, the solution describing magnetic field penetration in a superconductor predicts a penetration depth which is equal to $\lambda_L/\sqrt{2}$. In this section we will consider experimental data for YBCO, an extreme Type-II superconductor in which the coherence length $\xi$ is much smaller than $\lambda_L$ and the London theory with constant carrier density $n_s$ is expected to work well. We will show that the modified value $\lambda_L/\sqrt{2}$ is actually in good agreement with known estimates of the carrier density and of the ratio $m_{eff}/m$ between the effective and bare electron mass. This lends further support to the idea of a modified coupling for bosons, developed in this work based on the requirement of a linear coupling of the four-potential $A^\mu$ with a conserved current.

There are several ways of experimentally measuring the penetration depth and for YBCO they all agree on a value of approximately 150 nm (average on directions, since the single crystals are very asymmetric), extrapolated to $T=0$. The linear extrapolation is very accurate. See for example Fig.\ 4 in \cite{prozorov2000measurements}.   

According to the London formula $n_s=m_{eff}/(2\mu_0e^2\lambda^2)$, the value of $\lambda$ above would correspond to a carrier density $n_s=3.5\cdot 10^{27}(m_{eff}/m)$ m$^{-3}$. Using the modified formula this estimate is of course divided by 2, namely $n_s=1.75\cdot 10^{27}(m_{eff}/m)$ m$^{-3}$. It is possible to measure the carrier density independently, through the Hall effect. A recent measurement is reported in \cite{alcala2024tuning}. The carrier density found for high-quality YBCO with T$_c$ 90 K in low-resistance state, to which the data above for $\lambda$ refer, lies between 8 and 9 in units $10^{27}$ m$^{-3}$. This density value corresponds to nearly two carriers per chemical unit cell or 2/3 of a carrier per Cu atom.

A possible source of uncertainty in these estimates is the dependence of the Hall coefficient $R_H$ on the temperature. Close to T$_c$ and below it, $R_H$ has large variations, even becoming negative in a certain interval, when the applied field is small \cite{gob2000double}. This shows that the Hall conductivity in YBCO in that temperature range is, like for other cuprate superconductors, strongly influenced by the formation of holes with positive charge. Also for this reason, the authors of \cite{alcala2024tuning} have estimated the carrier density from $R_H$ at room temperature, where the dependence $R_H(T)$ is well stabilized [private communication by A.\ Palau, March 14, 2025].

Finally, the estimates for the ratio $m_{eff}/m$ in YBCO vary between 8 and 10 \cite{ivanov2003superconductivity}. This appears to give a better agreement with the modified formula for the penetration depth, compared to the traditional formula. 
 
Of course, more extensive checks need to be performed, possibly also using the modified Ginzburg-Landau equation of Sect.\ \ref{mod-gl}, which can describe a wider class of superconducting materials, provided the assumptions and approximations made in its derivation are satisfied.

\section{Conclusions}

As explained in detail in the Introduction, in this work we have proposed an alternative to the minimal e.m.\ coupling of matter fields which derives from local gauge invariance. The "new" principle we chose is valid also when local charge conservation is not strictly respected. We assumed a coupling term of the form $J^\mu A_\mu$, where $J^\mu$ is the N\"other current resulting from the global gauge invariance of the Lagrangian, supposed to be always true. This coupling is actually well known for fermionic fields and has a clear physical meaning: it gives a correct expression of the interaction energy as the work done by the e.m.\ field on the physical current of charged particles.

Our calculations with the extended Aharonov-Bohm electrodynamics (in which local charge conservation is not assumed from the start) show that for fermionic fields with the standard N\"other current $J^\mu=q\bar{\psi}\gamma^\mu\psi$ the usual formalism of QED is recovered and the coupling $J^\mu A_\mu$ coincides with that obtained from the minimal coupling recipe $\partial_\mu \psi \to \partial_\mu \psi + iqA_\mu \psi$.

This is not true, however, for scalar fields: the minimal coupling recipe does generate a coupling of the form $J^\mu A_\mu$, but $J^\mu$ does not coincide with the conserved N\"other current resulting from global invariance (eqs.\ \eqref{JHuang} and \eqref{Jscalar_conserved}). This fact is surprisingly overlooked in usual treatments of relativistic scalar field theories (see for ex.\ \cite{huang2010quantum,itzykson2012quantum}). We cannot tell yet if it could have any phenomenological consequences, as this issue lies outside the scope of the present work.

Applications of our alternative proposal to quantum field theory and particle physics are unlikely, because
giving up local gauge symmetry clearly has far-reaching consequences: Ward identities no longer hold, the photon propagator may cease to be transverse, and unphysical longitudinal modes can become dynamical and couple to matter. Also,
in standard QED, Ward identities ensure that the emission of soft photons factorizes universally and that infrared divergences cancel between real and virtual contributions (Bloch-Nordsieck and KLN theorems). In the Aharonov-Bohm framework for bosons, the field equations explicitly include non-conserved sources proportional to the gauge potential. This means that longitudinal photons can couple physically, breaking the universal factorization of soft emission.

Instead, we have considered the possible consequences for a system of non-relativistic bosons. By imposing the general requirement of a coupling $J^\mu A_\mu$, with $J^\mu$ a conserved N\"other current, we obtain the Schr\"odinger-like equation \eqref{sch-like}. Compared to the standard Schr\"odinger equation, it contains an additional diamagnetic term $\frac{q^2}{2m}|\mathbf{A}|^2\psi$ which doubles the diamagnetic term originating from the minimal coupling. Note that the current conserved by the Schr\"odinger-like equation is the same which is conserved by the standard Schr\"odinger equation, namely \eqref{conservedJ}. In this work we do not consider the possibility that this formal conservation is violated by effects of renormalization or wavefunction collapse.

It is clear that an anomalous factor 2 in the diamagnetic term, if applicable to an electron wavefunction, would be incompatible with the known phenomenology of Landau levels. In fact we obtain this anomalous factor only for bosons, and therefore in Sect.\ \ref{applicationSC} we focus on the case of Cooper pairs in superconductors, considering possible modifications   of the London equations. The modified theoretical estimate obtained for the magnetic penetration $\lambda$ in YBCO (an extreme Type-II superconductor well described by London theory)   appears to be compatible with experimental data, considering the uncertainties arising in the measurements not only from the value of $\lambda$ but also from those of the density carriers $n_s$ and of their effective mass.   Estimations of superconductor surface impedance and of critical fields in type-II superconductors using the present theory show that differences with the usual theory are fundamentally due also to the scaling of the penetration length and of the GL parameter $\kappa$. 
These are just preliminary checks which should be complemented with more systematic analysis for other superconductors.

\bibliography{sch} 
\bibliographystyle{ieeetr}

\end{document}